\documentclass{elsart}
\usepackage{natbib}

\usepackage{epsfig}

\usepackage{amssymb}

\begin{document}
\begin{frontmatter}
\title{A measure of centrality based on the network efficiency}

\author[label1]{Vito Latora} 
and \author[label2,label3]{Massimo Marchiori}

\address[label1] {Dipartimento di Fisica e Astronomia,
Universit\`a di Catania,\\
and INFN sezione di Catania, Corso Italia 57, 95129 Catania,
Italy}

\address[label2] {W3C and Lab. for Computer Science,
Massachusetts Institute of Technology, USA}

\address[label3]
{Dipartimento di Informatica, Universit\`a di Venezia, Italy}

\begin{abstract}
We introduce a new measure of centrality, the information 
centrality $C^I$, based on the concept of efficient propagation of 
information over the network. $C^I$ is defined for both valued 
and non-valued graphs, and applies to groups and classes as well 
as individuals. The new measure is illustrated and compared to 
the standard centrality measures by using a classic network 
data set.  
\end{abstract} 

\end{frontmatter}

\section{Introduction}
The idea of centrality was first applied to human communication 
by Bavelas \citep{bavelas48,bavelas50} who was interested 
in the characterization of the communication in small groups of 
people and assumed a relation between structural centrality 
and influence in group processes. Since then various measures of 
structural centrality have been proposed over the years to 
quantify the importance of an individual in a social network  
\citep{wasserman}. 
Most of the centrality measures are based on one of two quite 
different conceptual ideas and can be divided into two large 
classes. 
\\
The measures in the first class are based on the idea that  
the centrality of an individual in a network 
is related to how he is {\it near} to the other persons. 
The simplest and most straightforward way to quantify the 
individual centrality is therefore the degree of the individual, i.e. 
the number of its first neighbours.    
The most systematic elaboration of this concept is to be found in 
\citep{nieminen}.  
A degree-based measure of the individual centrality   
corresponds to the notion of how well connected the individual 
is within its {\it local} environment. 
The degree-based measure of centrality can be extended beyond 
first neighbours by considering the number of points that 
an individual can reach at distance two or three \citep{scott}. 
A {\it global} measure based on the concept of closeness 
was proposed in \citep{fre79} in terms of the distances among 
the various points. One of the simplest notion of closeness 
is that calculated from the sum of the geodesic distances from an 
individual to all the other points in the 
graph \citep{sabidussi}. 
\\
The second class of measures is based on the 
idea that central individuals stand {\it between} others on 
the path of communication \citep{bavelas48,anthonisse,fre77,fre79}. 
The betweenness of a point measures to what extent the point can 
play the role of intermediary in the interaction between the 
others. The simplest and most used measure of betweenness 
was proposed by Freeman  \citep{fre77,fre79} and is based on  
geodesic paths. In many real situations, however, communication 
does not travel through geodesic paths only. For such a reason 
two other measures of betweenness, 
the first based on all possible paths between a couple of 
points \citep{flow}, and the second based 
on random paths, \citep{newmanrandom} have been 
introduced more recently. 
\\
In this paper we propose a new measure of point centrality  
which is a combination of the two main ideas of centrality 
mentioned above. The new measure in fact is   
sensitive to how much an individual is close to the others 
and also to how much he stands between the others.  
The measure is named {\it information centrality} since is based 
on the concept of efficient propagation of information over 
the network \citep{lm2}. 
The information centrality of an individual is defined 
as the relative drop in the network efficiency caused by the 
removal of the individual from the network. 
In other words we measure how the communication over the network 
is affected by the deactivation of the individual. 
The information centrality is defined for both valued and 
non-valued graph, and naturally applies to group and classes as 
well as individuals. 
\\
The paper is organized as follows. In Section \ref{standard} 
we briefly review the most widely used measures of centrality, degree 
closeness and betweenness.  
In Section \ref{new} we introduce point and group information 
centrality, while in Section \ref{centralization} we discuss 
how the information centrality can be used to measure the 
centralization of the graph. 
Similarities and dissimilarities with respect to the standard 
measures are discussed and illustrated by means of simple examples in 
Section \ref{comparing} and Section \ref{primate}.

\section{The standard centrality measures} 
\label{standard}
We first review the three most commonly adopted 
measures of point centrality \citep{fre79}: 
the degree centrality $C^D$, the closeness centrality 
$C^C$, and the betweenness centrality $C^B$. Such measures 
of centrality imply three competing  
theories of how centrality might affect group processes, 
respectively centrality as activity, 
centrality as independence and centrality as control \citep{fre79}.  
We represent a social network as a 
{\it non-directed}, {\it non-valued} graph  ${\bf G}$, consisting 
of a set of $N$  points (vertices or nodes)  and a set of 
$K$ edges (or lines) connecting pairs of points. 
The points of the graph are the individuals, the actors of a 
social group and the lines represent the social links. 
The graph can be described by the so-called adjacency matrix, a 
$N \times N$ matrix whose entry $a_{ij}$ is $1$ if there 
is an edge between $i$ and $j$, and $0$ otherwise. 
The entries on the diagonal, values of $a_{ii}$, are 
undefined, and for convenience are set to be equal 
to 0. 
 
{\bf Degree Centrality} 
The simplest definition of point centrality is based on 
the idea that important points must be the 
most active, in the sense that they have the largest number of ties 
to other points in the graph. Thus a centrality measure for an actor 
$i$, is the degree of $i$, i.e. the number of points adjacent to $i$. 
Two points are said adjacent if they are linked by 
an edge. 
The degree centrality of $i$ can be defined as \citep{nieminen,fre79}: 
\begin{equation}
\label{DC}
C^D_i = \frac{k_i} {N-1} 
       = \frac{  \sum_{ j \in {\bf G}}   a_{ij}   } {N-1}
\end{equation}
where $k_i$ is the degree of point $i$. Since a given point $i$ 
can at most be adjacent to $N-1$ other points, $N-1$ is 
the normalization factor introduced to make the definition 
independent of the size of the network and to have 
$0 \le C^D_i \le 1$. 
The degree centrality focuses on the most visible actors in the 
network. An actor with a large degree is in direct contact to many 
other actors and being very visible is immediately recognized 
by others as a hub, a very active point and major channel of 
communication.

{\bf Closeness Centrality}
 The degree centrality is a measure of local centrality.  
A definition of actor centrality on a global scale is based on how close 
an actor is to all the other actors. In this case 
the idea is that an actor is central if it can 
quickly interact with all the others, not only with first neighbours.    
The simplest notion of closeness 
is based on the concept of minimum distance or  
geodesic $d_{ij}$, i.e. the minimum  number of edges  
traversed to get from $i$ to $j$.  
The closeness centrality of point $i$ is \citep{sabidussi,fre79,wasserman}: 
\begin{equation} 
\label{CC}
C^C_i = ({L_i})^{-1}  
       = \frac   {N-1}       {  \sum_{ j \in {\bf G}}   d_{ij}   }
\end{equation}
where $L_i$ is the average distance from actor $i$ to all the 
other actors and the normalization makes $0 \le C^C_i \le 1$. 
$C^C$ is to be used when measures based upon independence are 
desired \citep{fre79}. 
Such a measure is meaningful for connected graphs only, 
unless one assumes $d_{ij}$ equal to a finite value, 
for instance the maximum possible distance $N-1$, 
instead of $d_{ij}=+\infty$, when there is no path 
between two nodes $i$ and $j$. 
Such an assumption will be used in Section \ref{primate} 
to study a non-connected graph.

{\bf Betweenness Centrality}
Interactions between two non-adjacent points might depend 
on the other actors, especially on those on the paths between 
the two. Therefore points in the middle can have a strategic 
control and influence on the others. 
The important idea at the base of this centrality measure is that 
an actor is central if it lies between many of the actors. 
This concept can be simply quantified by assuming that 
the communication travels just along the geodesic.   
If $n_{jk}$ is the number of geodesics linking the two actors 
$j$ and $k$, and $n_{jk}(i)$ is the number of geodesics linking 
the two actors $j$ and $k$ that contain point $i$, 
the betweenness centrality of actor $i$ can be defined as 
\citep{anthonisse,fre77,fre79}: 
\begin{equation} 
\label{BC}
C^B_i  =  \frac{ {\sum_{{j < k\in {\bf G}}}}  n_{jk}(i)/ n_{jk}   } 
                    {(N-1)(N-2)}  
\end{equation}
In the double summation at the numerator, $j$ and $k$ must be different 
from $i$. 
Similarly to the other centrality measures $C^B_i$
takes on values between 0 and 1, and it reaches its maximum when 
actor $i$ falls on all geodesics.  
\\
There are several extensions to the original betweenness measure 
proposed by Freeman. 
In particular, in most of the cases the communication does 
not travel through geodesic paths only, and for this reason a 
more realistic betweenness measure should include non-geodesic as well 
as geodesic paths. Here we mention two other measures of betweenness 
that include contributions from non-geodesic paths: 
the flow betweenness and the random paths betweenness. 
We will disuss in detail the first one since it will be 
used in some of the examples of the following sections. 
The flow betweenness was introduced in 
\citep{flow} and is based on the concept of maximum flow. 
It is defined by assuming that each edge of 
the graph is like a pipe and can carry a unitary amount of flow 
(or an amount of flow equal to the edge's value in 
the extension to valued graphs). 
By considering a generic point $j$ as the source of flow 
and a generic target point $k$ as the target, it is possible 
to calculate the maximum possible flow from $j$ to $k$ by means 
of the min-cut, max-flow theorem \citep{ford}. 
In general it is expected that more than a single unit of flow 
is exchanged from $j$ to $k$ by making simultaneous 
use of the various possible paths.  
The flow betweenness centrality of point $i$ is defined from the 
amount of flow $m_{jk}(i)$  passing through $i$ when the maximum 
flow $m_{jk}$ is exchanged from $j$ to $k$, by the formula: 
\begin{equation} 
\label{FBC}
C^F_i  = 
 \frac{  \sum_{{j < k\in {\bf G}}} m_{jk}(i)  
      } 
      {    \sum_{{j < k\in {\bf G}}} m_{jk}     
      }  
\end{equation}
The second betweennes was introduced very 
recently in \citep{newmanrandom,newgirv2} and is 
based on random paths. It is suited for such cases in 
which a message moves from a source point $j$ to a target point 
$k$ without any global knowledge of the network, and therefore 
at each step chooses where to go at random from all the 
possibilities. The betweennes of point $i$ is equal to the number 
of times that the message passes through $i$ in its journey, averaged 
over a large number of trials of the random walk.  
\citep{newmanrandom}

\section{Point and group information centrality} 
\label{new}

In this section we introduce the information centrality, a new 
measure based on the concept of efficient propagation of information
over the network. The information centrality applies to points 
as well as to group/classes, and is defined both 
for valued and non-valued graph. For this reason 
we now consider a social network as a {\it non-directed} 
(however the extension to non-symmetric data -digraphs- do 
not present any special problem) {\it valued} graph 
${\bf G}$ of $N$ points and $K$ edges. A valued graph 
is a better description of a social system if the intensity 
of the social relations is a relevant ingredient that one 
wants to take into account. In fact the numerical value 
attached to each of the edges can be thought as a measure of 
the social proximity between two persons. Consequently 
the entries of the adjacency matrix $a_{ij}$ that describes 
${\bf G}$ are positive real numbers when there is an edge 
between $i$ and $j$, and $0$ otherwise.  
The most adopted convention is to consider the values 
$a_{ij}$ as proportional to the intensity of the social 
connection. An alternative, although equivalent description, 
that we adopt here, is to consider such numbers as 
inversely proportional to the intensity of the social connection;  
for instance $a_{ij}$ can be set to be 
equal to the inverse of the number of contacts 
between two individuals, or to the inverse of the amount of 
time they spend together. 
In our description the value of an edge can be imagined 
as a length associated to the edge: the 
stronger the intensity of the social link, the closer 
the two individuals are. In a valued graph,  
the shortest path length $d_{ij}$ between $i$ and $j$ 
is defined as the smallest sum of the
edges lengths throughout all the possible paths in the graph
from $i$ to $j$. 
When $a_{ij}=1$ for all existing edges, i.e.
in the particular case of a non-valued graph, 
$d_{ij}$ reduces to the minimum number of edges traversed to get 
from $i$ to $j$.
\\
The information centrality we are going to introduce 
is based on the following simple ideas: 
1) information in social networks travels in parallel in the 
sense that all the individuals exchange packets of information 
concurrently; 2) the importance of a point (group) is related to 
the ability of the network to respond to the deactivation 
of that point (group) from the network. 
In particular, we measure the network ability in propagating 
information among its points, before and after a certain 
point (group) is deactivated. 
\\
In order to measure how efficiently the points of the network 
${\bf G}$ exchange information, we use the {\em network efficiency\/} 
$E$, a quantity introduced in \citep{lm2,lm4}. 
The efficiency is a good measure of the performance of {\em parallel
systems\/}, i.e. when all the points in the graph 
concurrently exchange packets of information \citep{lm4}.   
Such a variable is based on the assumption 
that the information/communication in a network 
travels along the shortest routes and that the efficiency 
$\epsilon_{ij}$ in the communication 
between two points $i$ and $j$ is equal to the   
inverse of the shortest path lenght $d_{ij}$. 
The {\it efficiency} of ${\bf G}$ is the average of $\epsilon_{ij}$:   
\begin{equation} 
\label{efficiency}
E[{\bf G}]=
\frac{ {{\sum_{{i \ne j\in {\bf G}}}} \epsilon_{ij}}  } {N(N-1)}
          = \frac{1}{N(N-1)}
{\sum_{{i \ne j\in {\bf G}}} \frac{1}{d_{ij}}}
\end{equation}
and measures the mean flow-rate of information over ${\bf G}$. 
The quantity $E[{\bf G}]$ is perfectly defined in the 
case of non-connected graphs,  
in fact when there is no path between two points $i$ and $j$, 
we assume $d_{ij}=+\infty$ and consistently $\epsilon_{ij}=0$. 
For a non-valued graph $E$ varies in the range $[0,1]$.  We are now 
ready to define point and group information centrality. It is important 
to say that the same ideas can be applied to define the importance 
of the edges of the graph \citep{infra,next}

{\bf Point information centrality} 
The information centrality of a point $i$ is defined 
as the relative drop in the network 
efficiency caused by the removal from ${\bf G}$ 
of the edges incident in $i$:  
\begin{equation} 
\label{IC}
C^I_i  =  \frac{\Delta E}{E} = 
                \frac{E[{\bf G}] - E[{\bf G}^{\prime}_i]}{E[{\bf G}]}
\end{equation}
where by ${\bf G}^{\prime}_i $ we indicate a network with $N$ points 
and $K-k_i$ edges obtained by removing from ${\bf G}$ the edges 
incident in point $i$. 
The removal of some of the edges affects the communication 
between various points of the graph increasing the length of the  
shortest paths. Consequentely the efficiency of the new 
graph $E[{\bf G}^{\prime}_i]$ is smaller than $E[{\bf G}]$.  
The measure $C^I_i$ is normalized by definition to take 
values in the interval [0,1]. 
It is immediate to see that $C^I$ is somehow correlated to 
all the three stardard centrality measures:  
$C^D$ (formula \ref{DC}), $C^C$ (formula \ref{CC}), and 
$C^B$ (formula \ref{BC}). In fact, the 
information centrality of point $i$ depends on the degree of 
point $i$, since the efficiency $E[{\bf G}^{\prime}_i]$ 
is smaller if the number $k_i$ of edges removed from the 
original graph is larger. 
$C^I_i$ is correlated to $C^C_i$ since the efficiency of a 
graph is connected to $(\sum_i L_i)^{-1}$. Finally $C^I_i$, similarly to 
$C^B_i$, depends on the number of geodesics passing by $i$, 
but it also depends on the lenghts of the new geodesics,  
the alternative paths that are used as communication channels,  
once the point $i$ is deactivated. 
No information about the new shortest paths 
is contained in $C^B_i$, and in the other 
two standard measures.

{\bf Group information centrality}
Analogously to point centrality, 
the information centrality of a group of points ${\bf S}$ can be
defined as the relative drop in the network 
efficiency caused by the deactivation of the points in ${\bf S}$, 
i.e. by the removal from graph ${\bf G}$ of the edges incident in 
points belonging to $\bf S$:  
\begin{equation} 
\label{ICgroup}
C^I_{\bf S}  =  \frac{\Delta E}{E} = 
                \frac{E[{\bf G}] - E[{\bf G}^{\prime}_{\bf S}]}{E[{\bf G}]}
\end{equation}
Here by ${\bf G}^{\prime}_{\bf S}$
we indicate the network obtained by removing 
from ${\bf G}$ the edges incident in points belonging to $\bf S$. 
$C^I_{\bf S}$ is normalized to take values between 0 and 1.

\section{Graph Centralization} 
\label{centralization}
We have concentrated in so far on the question of the centrality 
of a point (and of a particular group of points) in the graph. 
But it is also possible to examine to which extent the whole 
graph has a centralized structure. 
In fact, related to the point centrality measures, is the idea of an 
overall index of centralization of a graph describing to 
which extent the graph is organized around its most central 
point. Indexes of graph centralization based on the standard 
measures of point centrality have been proposed over the 
years \citep{fre79}.  
Here we propose a measure of the graph centralization based on the 
information centrality. 
The two properties common to all the graph centralization 
indexes, no matter the point centrality measure upon which they are 
built on, are:    
1) graph centralization should measure to which extent the 
centrality of the most central point exceeds the centrality of 
the other points; 
2) graph centralization should be expressed as the ratio of that excess 
to its maximum possible value for a graph 
with the same number of points \citep{fre79}. 
We define a graph centralization based on the 
information centrality as: 
\begin{equation} 
C^I _{\bf G} =  \frac{ \sum_{i=1}^N  [C^I_{i^*} - C^I_i]} 
           {\max \sum_{i=1}^N  [C^I_{i^*} - C^I_i]}  = 
 \frac{ \sum_{i=1}^N  [C^I_{i^*} - C^I_i]}  {\frac{(N+1)(N-2)}{N+2} }
\end{equation}
where $i^*$ is the point with highest centrality. 
The normalization factor ${\frac{(N+1)(N-2)}{N+2} }$ is the maximum 
possible value of $\sum_{i=1}^N  [C^I_{i^*} - C^I_i]$, 
that is obtained for a star with $N$ points.

\section{Comparing $C^I$ to the other point centrality measures}
\label{comparing}
%
\begin{figure}
\begin{center}
\epsfig{figure=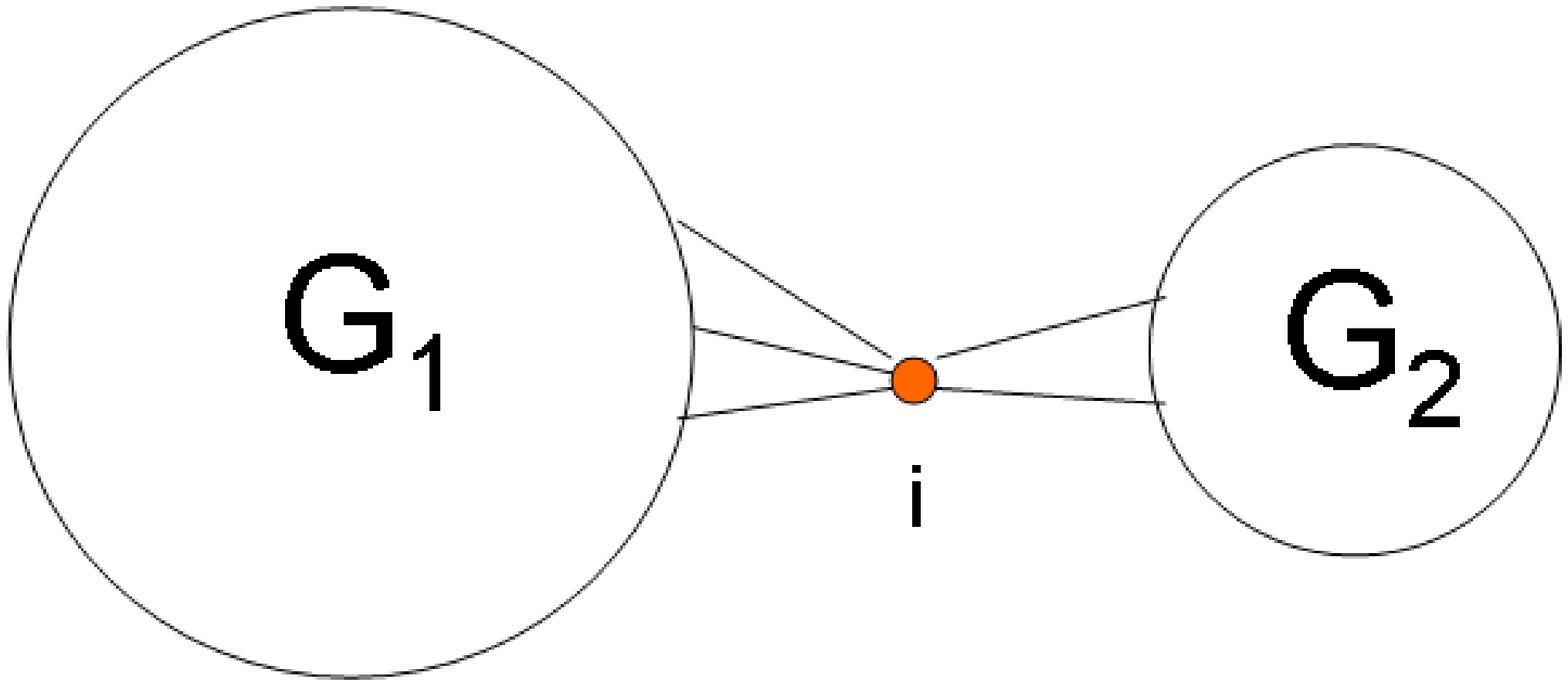,width=12truecm,angle=0}
\end{center}
\caption{A graph ${\bf G}$ composed by two subgraphs 
${\bf G_1}$ and ${\bf G_2}$ connected by node $i$ }
\label{fig0}
\end{figure}
%
%
The new measure of centrality we have introduced agrees with the 
three standard measures (degree, closeness, betweenness) on assignement 
of extremes. 
For instance it assignes the maximum importance 
to the central point of a star, and equal importance to the points 
of a complete graph. However the agreement breaks down 
between these extremes. 
Consider, for instance, the graph sketched in fig.~\ref{fig0}, which is 
composed by two main parts, graph ${\bf G_1}$ with $N_1$ points and 
graph ${\bf G_2}$ with $N_2$ points ($N_1 > N_2$), 
and by a single node $i$, connecting 
${\bf G_1}$ to ${\bf G_2}$. 
 For such simple example the information 
centrality contains some of the features of the betweenness 
(an actor is central if it lies between many of the actors).  
In fact the information centrality, similarly to 
the betweenness centrality, assigns the maximum importance to 
point $i$, which certainly plays an important 
role since it works as a bridge between ${\bf G_1}$ and ${\bf G_2}$. 
On the other hand it is very unlikely that the degree or the 
closeness centrality would attribute to point $i$ the highest 
score. 
The first, because ${\bf G_1}$ and ${\bf G_2}$ may  
contain points with degree larger than that of $i$. 
The second, because the point with smallest distance to 
all the other points 
will probably be in ${\bf G_1}$, especially if we assume 
$N_1 \gg N_2$.  
%
\begin{figure}
\begin{center}
\epsfig{figure=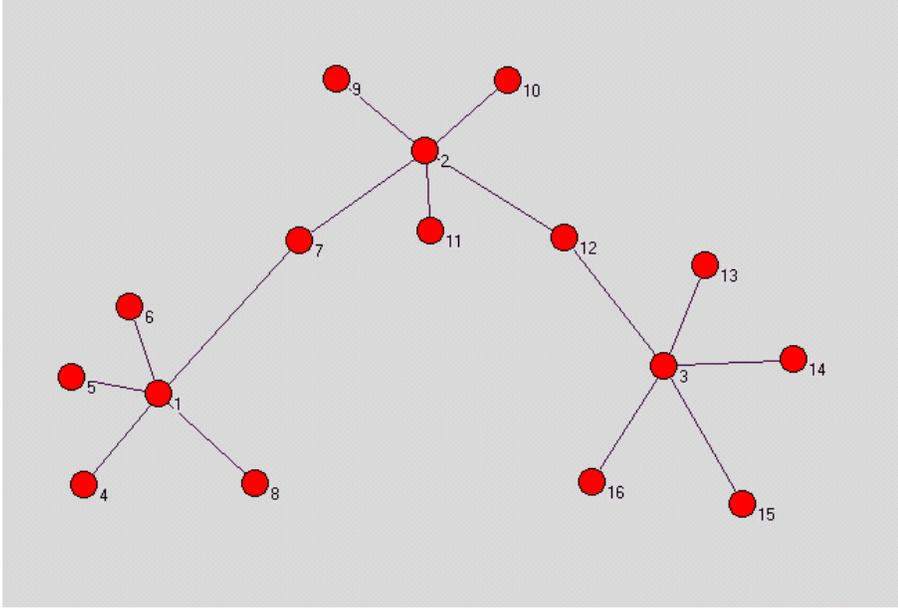,width=12truecm,angle=0}
\end{center}
\caption{A non-directed non-valued tree 
with N=16 points, a simple case constructed 
to compare the new measure of centrality we have 
introduced with the three standard measures: degree, 
closeness and betweenness. 
}
\label{fig1}
\end{figure}
%
%
%
We will now illustrate similarities and dissimilarities with 
the three standard measures by using as an example 
a non-directed non-valued graph constructed ad hoc. 
The graph considered, drawn in fig.\ref{fig1}, is a tree with 
N=16 points and K=N-1. The four centrality scores are reported 
in tab.\ref{table1}. The points are ordered in decreasing 
order of $C^I$. 
%
\begin{table}
\caption{The point centrality $C^I$ is compared to the standard 
centrality measures $C^D$, $C^C$, and $C^B$ for the graph in  
fig.\ref{fig1}. 
The points are ordered according to $C^I$. 
\label{table1}} 
\begin{tabular}{l||l||l|l|l}
  point & $C^I$  &  $C^D$   &   $C^C$  &    $C^B$  \\
\hline
 2 &   0.591&   0.333&   0.455&   0.714\\
 1 &   0.444&   0.333&   0.349&   0.476\\
 3 &   0.444&   0.333&   0.349&   0.476\\
 7 &   0.389&   0.133&   0.405&   0.476\\
12 &   0.389&   0.133&   0.405&   0.476\\
 9 &   0.116&   0.067&   0.319&   0.000\\
10 &   0.116&   0.067&   0.319&   0.000\\
11 &   0.116&   0.067&   0.319&   0.000\\
 4 &   0.106&   0.067&   0.263&   0.000\\
 5 &   0.106&   0.067&   0.263&   0.000\\
 6 &   0.106&   0.067&   0.263&   0.000\\
 8 &   0.106&   0.067&   0.263&   0.000\\
13 &   0.106&   0.067&   0.263&   0.000\\
14 &   0.106&   0.067&   0.263&   0.000\\
15 &   0.106&   0.067&   0.263&   0.000\\
16 &   0.106&   0.067&   0.263&   0.000\\
\end{tabular}
\end{table}
Although the four measures show a certain overall agreement, 
for instance they all attribute the highest centrality to 
point $2$, there are some differences worth of 
noting. 
The information centrality assignes the top score to point 2, 
second score to points 1,3, third score to points 7,12. 
But it also distinguish point 9,10,11 (fourth score) from the 
remaining points. The only other measure that operates such 
a distinction is $C^C$ which, on the other hand, assignes 
the second score to points 7,12 and the third score to 
points 1,3 inverting the result of $C^I$. 
Neither the degree centrality $C^D$ nor 
the betweenness centrality $C^B$ have the resolution of 
$C^I$ and $C^C$. In fact $C^D$
assignes the top score to three 
points, namely points 1,2,3 all having five neighbours,  
and the second score to points 7,12 both with 2 neighbours. 
While $C^B$ assignes the top score to point 2 and the 
second score to points 1,3,7,12. Both $C^D$ and $C^B$ does not 
distinguish points 9, 10, 11 from the remaining 
points: 4,5,6,8,13,14,15,16.  
From this simple example $C^I$ results as having, 
together with $C^C$, the best resolution. On the other hand, as 
we have seen in the example of fig.~\ref{fig0}, 
$C^I$ contains some of the features 
of $C^B$. And in the next session we will show that 
in some cases  $C^I$ can be strongly correlated to $C^D$.

\section{Applications to the primate data} 
\label{primate}
\begin{figure}
\begin{center}
\epsfig{figure=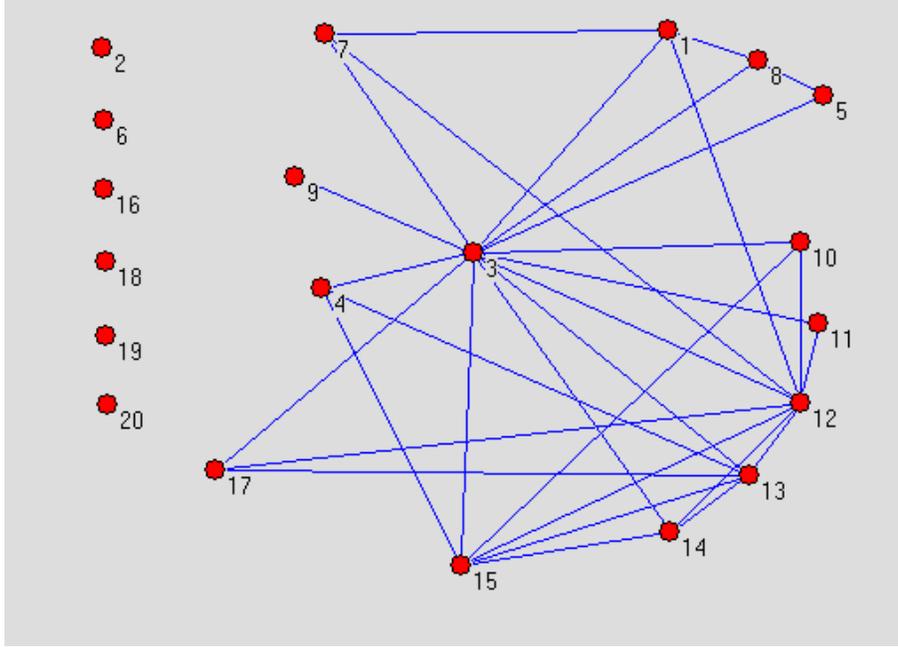,width=12truecm,angle=0}
\end{center}
\caption{The graph of the interactions amongst a group of 20 
monkeys. The dataset collected by Linda Wolfe \citep{ucinet,groups} 
contains also information on the sex and age of each animal (see table 
\ref{table2}) 
}
\label{fig2}
\end{figure}

\begin{figure}
\begin{center}
\epsfig{figure=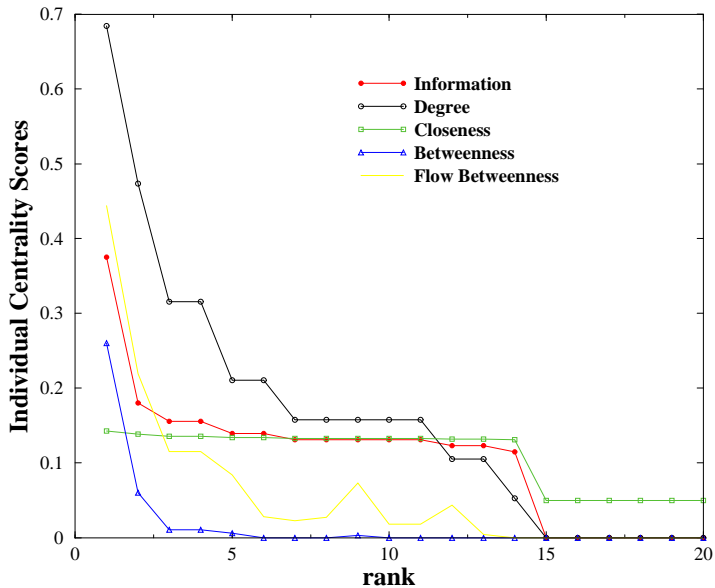,width=8truecm,angle=270}
\end{center}
\caption{Individual centrality score for each of the 20 points of 
the graph of interactions amongst monkeys. The points are ordered 
according to their value of $C^I$ (see table \ref{table2}). 
}
\label{fig3}
\end{figure}

\begin{figure}
\begin{center}
\epsfig{figure=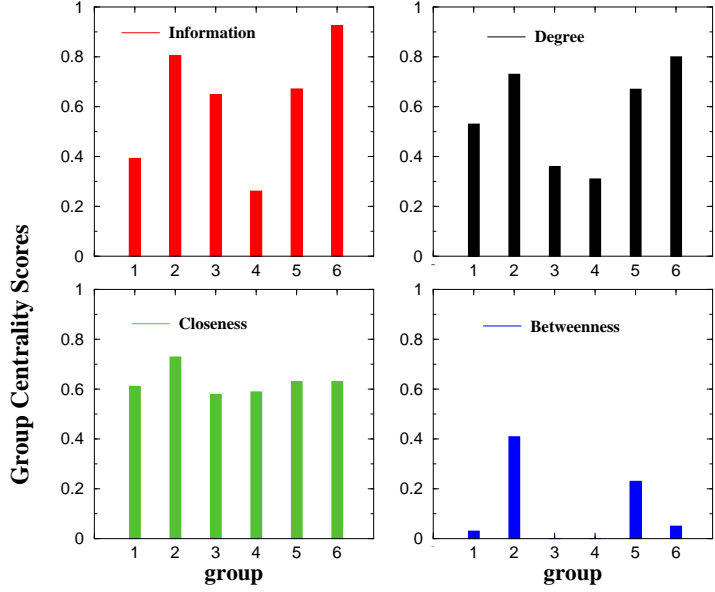,width=8truecm,angle=270}
\end{center}
\caption{Centrality score for each of the six groups 
considered, namely 1(age 14-16), 2(age 10-13), 3(age 7-9), 
4(age 4-5), 5 (males), 6 (females),  
and the four centrality measures}
\label{fig4}
\end{figure}

In this section we study a classical data set, the 
primate data collected by Linda Wolfe \citep{ucinet,groups}, 
recording 3 months of interactions amongst a group of 
20 monkeys, where interactions were defined as the joint presence 
at the river. The resulting non-directed non-valued 
graph is represented in fig.\ref{fig2}. The graph consists 
of 6 isolated points and a connected component of 14 points.  
The dataset also contains information on the sex and the age 
of each animal as reported in tab.\ref{table2}. 
%
%
Such a graph has been studied in \citep{groups}, where  
the standard point centrality measures of degree, closeness and 
betweenness have been generalized to apply to groups as well as 
individuals. For such a particular dataset we can therefore 
compare the measure we have introduced to the standard 
measures of point centrality and also of group centrality.  
\\
In table \ref{table2} we report the point centrality scores obtained 
for each monkey, respectively $C^I$, $C^D$, $C^C$ and $C^B$. 
The flow betweenness centrality $C^F$ ~\citep{flow}
is also considered. As discussed in Section \ref{standard} 
the flow betweenness is not based 
on geodesic paths as in $C^B$, but on all the independent 
paths between all pairs of points in the graph. 
Age and sex of each monkey are also reported in table.    
Monkey 3 results the most central according to all the centrality 
indexes considered. 
Again all the centrality measures 
considered assignes the second rank to monkey 12 and the 
third rank to monkeys 13 and 15. 
The six isolated monkeys are the least central points according to
$C^I$, $C^D$ and $C^C$. Notice that for any of these six points  
$C^C$ is equal to 0.05 and not to zero, since it is assumed 
that $d_{ij} = N~~\forall j$  and 
therefore $C_i^C=L_i^{-1}= (N-1) / (N-1)N $. On the other hand,  
the betweenness centrality $C^B$ assignes a zero score to fourteen 
points,  the six isolated monkeys and other eight monkeys, namely 
4,5,7,9,10,11,14,17. The latter points, although having a degree 
equal or larger than one 
- for instance monkey 14 has four neighbours, while monkeys 
4,7,10 and 17 have three neighbours each - 
do not play any role in the communication 
between couples of points, in the sense that are not present 
in the shortest paths between couples of points. 
Of course such a result is a consequence of the assumption that 
communication between couples of point takes only the shortest path 
In fact, by considering the flow betweenness, only seven points have 
a zero score, the six isolated points and monkey 9. 
%
\begin{table}
\caption{Individual centralities: for each monkey we report 
age and sex group, the information centrality $C^I$ and 
the three standard centrality measures $C^D$, $C^C$ and  $C^B$. 
The flow betweenness centrality $C^F$ is also reported in 
the last column.  
\label{table2}} 
\begin{tabular}{l|l|l||l||l|l|l|l|l|l}
  Monkey & Age group & Sex & $C^I$  &  $C^D$   &     $C^C$    &    $C^B$   &   $C^F$\\
\hline
 1 & 14-16 & Male & 0.1393&  0.2105&  0.1338&  0.0058&  0.0841 \\
 2 & 10-13 & Male & 0.0000&  0.0000&  0.0500&  0.0000&  0.0000 \\
 3 & 10-13 & Male & 0.3751&  0.6842&  0.1429&  0.2602&  0.4439 \\
 4 &  7-9  & Male & 0.1311&  0.1579&  0.1329&  0.0000&  0.0228 \\
 5 &  7-9  & Male & 0.1230&  0.1053&  0.1319&  0.0000&  0.0437 \\
 6 & 14-16 & Female & 0.0000&  0.0000&  0.0500&  0.0000&  0.0000 \\
 7 &  4-5  & Female & 0.1311&  0.1579&  0.1329&  0.0000&  0.0274 \\
 8 & 10-13 & Female & 0.1311&  0.1579&  0.1329&  0.0029&  0.0731 \\
 9 & 7-9   & Female & 0.1148&  0.0526&  0.1310&  0.0000&  0.0000 \\
10 &  7-9  & Female& 0.1311&  0.1579&  0.1329&  0.0000&  0.0183 \\
11 & 14-16 & Female & 0.1230&  0.1053&  0.1319&  0.0000&  0.0044 \\
12 & 10-13 & Female & 0.1803&  0.4737&  0.1387&  0.0604&  0.2195 \\
13 & 14-16 & Female & 0.1557&  0.3158&  0.1357&  0.0107&  0.1154 \\
14 &  4-5  & Female & 0.1393&  0.2105&  0.1338&  0.0000&  0.0280 \\
15 &  7-9  & Female & 0.1557&  0.3158&  0.1357&  0.0107&  0.1154 \\
16 & 10-13 & Female & 0.0000&  0.0000&  0.0500&  0.0000&  0.0000 \\
17 &  7-9  & Female & 0.1311&  0.1579&  0.1329&  0.0000&  0.0183 \\
18 &  4-5  & Female & 0.0000&  0.0000&  0.0500&  0.0000&  0.0000 \\
19 & 14-16 & Female & 0.0000&  0.0000&  0.0500&  0.0000&  0.0000 \\
20 & 4-5   & Female & 0.0000&  0.0000&  0.0500&  0.0000&  0.0000 \\
\end{tabular}
\end{table}
For the dataset considered, the ranking of the 20 points 
produced by $C^I$, $C^D$ and $C^C$ is the same. 
Nevertheless the normalized values of these measures 
are different as reported in table \ref{table2} and as can be 
seen in Fig.~\ref{fig3}, where we plot the centrality score 
for each of the 20 points. 
The points are ordered as a function of their score according 
to $C^I$. The behavior of $C^I$, $C^D$ and $C^C$ is similar although  
for the first five points the normalized values of 
$C^I$ are smaller than $C^D$ and larger than $C^C$. 
For instance the first point in the rank, namely monkey 3, has 
$C^D=0.6842$, $C^I=0.3751$ and $C^C=0.1429$. 
The two betweenness measures show some discrepancy with respect 
to the other measures. This is particularly evident 
in the figure for the flow betweenness: the two peaks 
at rank 9 and rank 12, corresponding respectively to 
point 8 and point 5, indicate that such two monkeys 
have, according to the flow betweenness, a rank larger that that 
assigned according to $C^I$.
%
%
\\
We now consider the six different groups studied 
in \citep{groups}: four formed by age and 
two formed by sex. 
Group 1 contains the five monkeys having age 14-16,   
group 2 the five monkeys having age 10-13, 
group 3 the six monkeys having age 7-9 and 
group 4 the four monkeys having age 4-5. 
Group 5 is made by the five females, while group 6 
is made by the fifteen males. 
%
%
As illustrated in Fig.~\ref{fig4}, among the age groups 
the most central one is  the 10-13 years old (group 2), 
according to all the four measures. This is the group containing monkey 
3, who is the most central point also as an individual. 
The four age groups in decreasing order of importance are:  
2, 3, 1, 4 for the information centrality, 
2, 1, 3, 4 for the degree centrality, 
2, 1, 4, 3 for the closeness centrality and 
2, 1, 3-4  for the betweenness centrality which assigns a score 
equal to zero to the two youngest groups. 
\\
The information centrality is the only measure assigning 
the second position to the 7-9 years old, while the other three 
measures assign the second position to the 14-16 years old. 
The information centrality assigns last position to group 4 
(age 4-5), similarly to the degree centrality and to the betwenness 
centrality. 
\\
Among the sex groups the most central one is the male group (which 
is also the largest one) for 
both information and degree centrality. 
The situation is inverted according to the betweenness centrality, 
while the closeness centrality attributes the same score to the 
two groups. 
\\
In addition to groups formed a priori, like a team in a company 
or the division of the individuals according to age 
or sex, like the one we have considered above,  
the centrality measure we have proposed in this paper can be 
applied to set of individuals identified by cohesive subgroups 
techniques such as cliques, n-cliques, 
k-plexex, lambda sets etc. \citep{wasserman}). 
\\
Another possibility is to use the centrality measure as a criterion 
to forming groups: we are working on an algorithm 
to finding the groups inside a given graph based on the concept 
of graph efficiency and information centrality \citep{next}.  

\section{Conclusions}
\label{conclusions}
In this paper we have briefly reviewed the standard measures of 
centrality proposed for social networks and we have introduced a new 
measure of centrality, the information centrality $C^I$, that 
is based on the concept of efficient propagation of 
information over the network. $C^I$ is defined for both valued and 
non-valued graphs, and applies to groups as well 
as individuals. The groups can be either a set of individuals formed a 
priori, such as a team in a company or a group of individuals chosen
according to some attribute (age, sex, income), 
or a set of individuals identified by cohesive subgroups 
methods or by positional analysis method. 
We have illustrated similarities and dissimilarities with respect 
to the three standard measures of degree, closeness and 
betweenness in two non-directed non-valued graphs. 
\\
It remains to be seen if, in the light of further empirical work, 
the new measure can be more appropriate than the others 
in some applications.

We thank S.P. Borgatti and M.G. Everett for providing us with the 
database of primate interactions, and P. Crucitti and G. Politi 
for a critical reading of the manuscript. 
%
\bigskip
\noindent

\end{document}